\documentclass[aps,pre,reprint,superscriptaddress,longbibliography,floatfix]{revtex4-2}
\usepackage{graphicx}
\usepackage{amsmath,amssymb,amsfonts}
\usepackage{mathrsfs}
\usepackage{xcolor}
\usepackage{etoolbox}
\makeatletter
\patchcmd{\@bibdataout@aps}{pages="0"}{pages="1"}{}{\errmessage{Bibliography page-range setup failed}}
\makeatother
\usepackage[colorlinks=true,citecolor=blue,linkcolor=blue,urlcolor=blue]{hyperref}
\begin{document}
\title{Phase transitions in the simplicial Ising model on hypergraphs}
\author{Gangmin Son}
\email{gmson102@gmail.com}
\affiliation{School of Computational Sciences, Korea Institute for Advanced Study, Seoul 02455, Korea}
\author{Deok-Sun Lee}
\email{deoksunlee@kias.re.kr}
\affiliation{School of Computational Sciences, Korea Institute for Advanced Study, Seoul 02455, Korea}
\affiliation{Center for AI and Natural Sciences, Korea Institute for Advanced Study, Seoul 02455, Korea}
\author{K.-I. Goh}
\email{kgoh@korea.ac.kr}
\affiliation{Department of Physics, Korea University, Seoul 02841, Korea}
\begin{abstract}
Complex higher-order interactions can reshape higher-order dynamics in complex systems. Systematic understanding of the repertoire of phase transitions that higher-order interactions can trigger remains incomplete. Here we study the phase transitions in the simplicial Ising model on hypergraphs, in which the energy within each hyperedge is lowered only when all the member spins are unanimously aligned. We show that its Hamiltonian is equivalent to a weighted sum of lower-order interactions and use the Landau free energy to identify diverse phase transitions depending on the hyperedge sizes. For $q$-uniform hypergraphs, the nature of transitions shifts from continuous to discontinuous at the tricritical point $q=4$. When both pairwise edges and hyperedges of size $q>2$ coexist, diverse scenarios emerge, including mixed-order and double transitions, for $q>8$. We further adopt the Bethe--Peierls method to investigate the interplay between pairwise and higher-order interactions in achieving global magnetization, illuminating the multiscale nature of double transitions.
\end{abstract}
\maketitle

\section{Introduction}

Pairwise interactions have widely been assumed in modeling the dynamics of complex systems; however, this assumption may fail when group or higher-order interactions (HOIs) are relevant~\cite{battiston2020networks,bianconi2021higher,boccaletti2023structure}.
Specifically, it has been revealed that HOIs affect collective phenomena including contagion dynamics~\cite{iacopini2019simplicial,st2021universal,kim2024higher,ferraz2024contagion}, percolation~\cite{bianconi2024theory,lee2023k,bianconi2024nature}, synchronization~\cite{skardal2019abrupt,millan2020explosive}, diffusion~\cite{carletti2020random,di2024dynamical}, evolutionary game~\cite{civilini2021evolutionary,civilini2024explosive}, and opinion dynamics~\cite{noonan2021dynamics,horstmeyer2020adaptive,papanikolaou2022consensus,kim2025competition}.
In particular, HOIs provide a natural route to abrupt or ``explosive'' phenomena~\cite{battiston2021physics,kuehn2021universal}.
For example, the simplicial contagion model~\cite{iacopini2019simplicial} adopts the so-called ``simplicial rule'' that an individual in a group can be infected if all the other group members are unanimously infected, which gives rise to discontinuous transitions in global social adoption~\cite{kim2023contagion,burgio2024triadic}.

\begin{figure}[b]
    \centering
    \includegraphics[width=.99\linewidth]{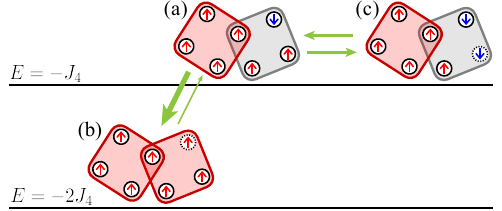}
    \caption{
    Simplicial Ising model (SIM). The SIM on a 4-uniform hypergraph with two hyperedges is schematically illustrated. Three configurations (a--c) are displayed along the energy level $E$ based on Eq.~(\ref{eqn:Hamiltonian}). The green arrows represent single-flip transitions with their widths illustrating transition probabilities at a finite temperature. The energy for a hyperedge changes by a single spin flip only between a state where all spins are aligned except for one (a) and a unanimous state (b). Otherwise, a spin-flip does not lead to an energy difference, as in the transitions between (a) and (c).
    }
    \label{fig1}
\end{figure}

Despite a vast body of research on nonequilibrium dynamics in HOI networks, the Ising model, an archetypal equilibrium system in statistical physics, has not been thoroughly explored in the context of HOIs.
Extensions of the Ising model have repeatedly been a stepping stone for advances across diverse domains, from spin glass~\cite{edwards1975theory} to, more recently, active matter~\cite{solon2013revisiting}.
In particular, with the advent of complex network theory~\cite{dorogovtsev2008critical}, the Ising model has been a foundational framework for understanding the role of structural properties in phase transitions and critical phenomena~\cite{dorogovtsev2002ising,leone2002ferromagnetic,bianconi2002mean}. Therefore, extending the Ising model to HOI networks may pave the way for integrating recent advances in HOIs. Recently, an Ising model with the Hamiltonian formulated under the simplicial rule, named the simplicial Ising model (SIM), was introduced in Ref.~\cite{robiglio2025synergistic}.
In the model, the pairwise coupling of two spins is extended to the exclusive preference for unanimity among multiple spins (see Fig.~\ref{fig1}), ensuring that the SIM preserves the $\mathbb{Z}_2$ symmetry of the Ising model.
Similar unanimous decision making may play a role in social dynamics~\cite{castellano2009statistical,budrikis2024100,stauffer2008social} such as jury decisions~\cite{lee2015statistical}.
However, many of its properties including phase transitions remain to be understood, warranting further investigation.

In this paper, we explore the phase transitions in the SIM.
Using the Bragg--Williams mean-field theory, we show that HOIs, here referring to the simplicial (unanimity) interactions, yield diverse scenarios of phase transitions depending on the sizes of hyperedges.
First, we consider $q$-uniform hypergraphs in which all hyperedges have the same size $q$.
We show that discontinuous transitions occur when $q>4$ and that the transition temperatures as a function of $q$ can be nonmonotonic due to the ambivalent influence of group size $q$ on the lower-order energetic terms in the free energy.
Furthermore, the coexistence of pairwise interactions and HOIs can give rise to mixed-order and double transitions, especially when the size $q$ of HOIs is $q>8$. We identify the distinct but synergistic roles of pairwise interactions and HOIs in the emergence of double transitions by using the Bethe--Peierls method.

\section{Results}
\begin{figure}[t]
    \centering
    \includegraphics[width=.99\linewidth]{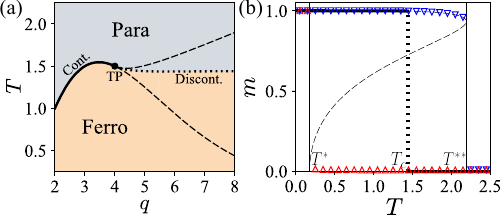}
    \caption{Phase transitions of the SIM on $q$-uniform hypergraphs. (a) Phase diagram in the $(q,T)$-space. The solid (dotted) line corresponds to continuous (discontinuous) transitions at the thermodynamic transition temperature $T_c$. The circle indicates the tricritical point (TP) at $q=4$. The two dashed lines represent the limits of metastability on heating (upper, $T^{**}$) and cooling (lower, $T^{*}$). (b) Order parameter $m$ as a function of $T$ for $q=10$. The solid line represents the location of the global minimum of the free energy, while the dotted line denotes $T_c$ across which the global minimum changes discontinuously. The dashed line corresponds to the solutions of $\partial f(m) / \partial m = 0$, partially overlapping with the solid line. The red upward (blue downward) triangles are the results of Monte Carlo simulations initialized from fully ordered (random) spin configurations on fully connected hypergraphs with $N=10^4$ nodes (see Methods for details).}
    \label{fig2}
\end{figure}

\begin{figure*}[t]
    \centering
    \includegraphics[width=.99\linewidth]{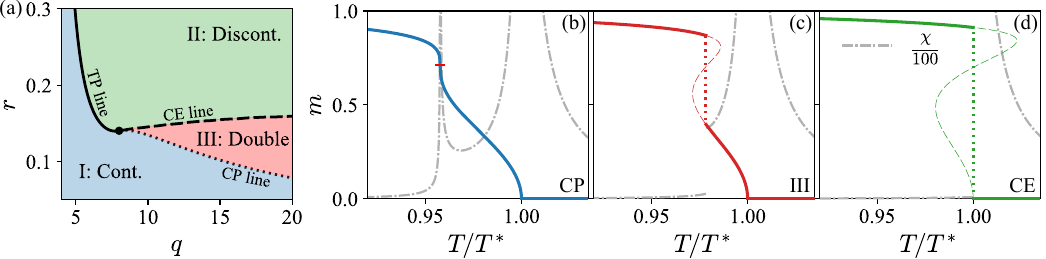}
    \caption{
    Phase transitions in $(2,q)$-hypergraphs. (a) Phase diagram on the $(q,r)$-space. Regimes I, II, and III are divided by the TP, CP, and CE lines. The abbreviations are: TP: tricritical point; CP: critical point; CE: critical endpoint (see the main text for their descriptions). The circle indicates the ``special'' TP at $q=8$. In Regime I, a continuous transition occurs. In Regime II, a discontinuous transition occurs. In Regime III, a continuous and a discontinuous transition occur successively. (b-d) Diverse phase transition scenarios in $(2,12)$-hypergraphs for $r=r_{\text{CP}}$ (b), $r_{\text{CP}}<r<r_{\text{CE}}$ (c), and $r = r_{\text{CE}}$ (d). The colored lines illustrate the order parameter. The solid line corresponds to the location of the global minimum of the free energy, while the dotted line denotes the discontinuous transition. The dashed line reflects the solutions of $\partial f(m) / \partial m = 0$, partially overlapping with the solid line. The gray dashed-dotted line corresponds to the susceptibility $\chi$.
    }
    \label{fig3}
\end{figure*}

\subsection{Model and mean-field theory}
The simplicial Ising model (SIM)~\cite{robiglio2025synergistic} is defined on a hypergraph $(\mathcal{V},\mathcal{E})$ with the set of nodes $\mathcal{V}=\{1 \ldots N\}$ and the set of hyperedges $\mathcal{E}$. Each node $i$ is assigned a spin $S_i\in\{\pm 1\}$ with $i\in\mathcal{V}$.
We consider the Hamiltonian $\mathcal{H}$ of the SIM given by
\begin{align}
    \label{eqn:Hamiltonian}
    \mathcal{H}
    &= -\sum_{ e \in \mathcal{E}} J_{|e|}  \delta_{\{S_i\}_{i \in e}} - H\sum_{i\in \mathcal{V}}{S_i},
\end{align}
where the ferromagnetic coupling constant $J_{|e|}>0$ is given by a function of the size (cardinality) $|e|$ of hyperedge $e$, and $H$ is an external field. 
The simplicial interaction of $q=|e|$ spins in a hyperedge $e$ is represented by
\begin{align}
    \delta_{\{S_1, \ldots, S_{q}\}}
    &= \begin{cases}
    1 & \text{if } S_1 = \ldots = S_{q},\\
    0 & \text{otherwise},
    \end{cases}\label{eqn:delta_def1}
\end{align}
which we will refer to as the higher-order Kronecker delta (HOKD).
Therefore, the energy for a hyperedge of size $q$ is $-J_q$  only in the two unanimous states, where all spins are aligned to either $S_i=1$ or $S_i=-1$, and it is $0$ in all the remaining $2^q-2$ nonunanimous states (see Fig.~\ref{fig1}).

One can represent the HOKD  as the sum of two products corresponding to the two unanimous states
\begin{equation}
    \delta_{\{S_i\}_{i \in e}}
    = \prod_{i \in e}{\left( \frac{1+S_i}{2} \right)} + \prod_{i\in e}{\left( \frac{1-S_i}{2} \right)}.
    \label{eqn:delta_def2}
\end{equation}
Expanding Eq.~(\ref{eqn:delta_def2}) as a power series of $S$'s, one finds that the energy of a hyperedge can be represented by the sum of the products of all spins in every even-sized subset of the hyperedge and therefore the Hamiltonian in Eq.~(\ref{eqn:Hamiltonian}) can be represented as  
\begin{equation}
\mathcal{H}=   -\sum_{e\in \mathcal{E}} \frac{J_{|e|}}{2^{|e|-1}}\sum_{\substack{e' \subset e \\ |e'| \text{ even}}}\prod_{i \in e'}{S_i} -H\sum_{i\in{\cal V}}S_i
    \label{eqn:Hamiltonian2}
\end{equation}
up to a constant.
This includes the Hamiltonians for the $p$-spin Ising model~\cite{gross1984simplest,franz2001ferromagnet} applied to all even-sized subsets of each original hyperedge, reminiscent of a simplicial complex~\cite{bianconi2021higher}, where all subsets, i.e., faces, of a simplex, also belong to the complex, justifying the naming of the SIM.
Note that Eq.~(\ref{eqn:Hamiltonian2}) on pairwise networks is equal to the Hamiltonian of the conventional Ising model.

To explore the phase transitions of the SIM, we first adopt the Bragg--Williams mean-field theory~(see Methods for details), in which every node is assumed to be subject to a single effective field represented in terms of the global magnetization $m=\sum_{i=1}^{N} \langle S_i \rangle / N$ with $\langle \cdot \rangle$ representing the ensemble average.
The Landau free energy density $f(m)$ is given by 
\begin{align}\label{eqn:free_energy}
    f(m) &=
     -\sum_{q=2}^{\infty} \rho_{q} J_{q} \left[ \left( \frac{1+m}{2} \right)^{q} + \left( \frac{1-m}{2} \right)^{q} \right] -Hm\nonumber\\
    &+ T \left( \frac{1+m}{2} \ln{\frac{1+m}{2}} + \frac{1-m}{2} \ln{\frac{1-m}{2}} \right),
\end{align}
where $\rho_{q} \equiv |\mathcal{E}_{q}|/N$ is the density of the size-$q$ hyperedges and $T$ is the temperature. The second line corresponds to $-Ts(m)$ with the entropy density, $s(m)$.   Note that, for pairwise networks, i.e., $\rho_q=0$ for all $q>2$,  $f(m) = {\rho_2 J_2 \over 2} (1+m^2) - Ts(m)$.

To understand analytically the transition from the paramagnetic ($m=0$) to the ferromagnetic ($m>0$) phase, one can expand $f(m)$ around $m=0$ as
\begin{align}
    f(m) - f(0) = \sum_{n=1}^{\infty} C_{n} m^{n},
    \label{eqn:free_energy2}
\end{align}
where the coefficients $C_n$ are given by
\begin{align}
    C_n = - \sum_{q \geq n}^{\infty} \frac{\rho_{q} J_{q}}{2^{q - 1}} {{q}\choose{n}} + \frac{1}{n(n-1)} T
    \label{eqn:landau_coeff}
\end{align}
for even $n$ and $C_n=0$ for odd $n$ except for $C_1 = - H$. Like Eq.~(\ref{eqn:Hamiltonian2}), this expansion reveals how the simplicial interaction of spins within a hyperedge of size $q$ influences all the lower-order energetic contributions ($n\leq q$) to the free energy while leaving the entropic term unchanged. The finding that the HOKD is equivalent to a sum of spin products in Eq.~(\ref{eqn:Hamiltonian2}) and the following explicit formulation of the free energy in Eqs.~(\ref{eqn:free_energy}--\ref{eqn:landau_coeff}) are the key contribution of this paper.

\subsection{Single-size groups}
Let us first consider the SIM on $q$-uniform hypergraphs in which all hyperedges are of the same size $q$.
Setting $\rho_q J_q = 1$ and $H=0$ and tracing the global and local minima of $f(m)$, we obtain the phase diagram in the $(q,T)$-space as shown in Fig.~\ref{fig2}(a).
Discontinuous transitions occur for $q > 4$ with  the tricritical point (TP) at $(q, T) = (4,\frac{3}{2})$ obtained from $C_2 = C_4 = 0$, while $C_6>0$.
The critical exponent $\beta$ at $q=4$ is $\beta_{\rm TP} ={1 \over 4}$ rather than $1 \over 2$ at the continuous transitions for $q<4$.
That the threshold is at $q=4$ is hinted at by Eq.~(4): for $q=3$, the higher-order Kronecker delta reduces to pairwise terms, whereas a genuine higher-order coupling first appears at $q=4$.

When $q > 4$, in addition to the thermodynamic transition temperature $T_c$, the boundaries of metastability on heating, $T^{**}$, and cooling, $T^{*}$, are also shown in Fig.~\ref{fig2}(a).
As $q$ increases, the metastable region $T^{**} - T^{*}$ broadens, and $T^{*} \to 0$ and $T^{**} \to \infty$ as $q\to\infty$ (see Methods for details).
In addition, the jump size $\Delta m$ goes to $\Delta m \to 1$ so that the value of $m$ can be $1$ or $0$ depending on the initial configuration. Figure~\ref{fig2}(b) shows the case of $q=10$ with $\Delta m \approx 0.94$; Monte Carlo simulations on fully connected $10$-uniform hypergraphs by the Metropolis algorithm agree with the predictions of the mean-field theory~(see Methods for details).

The transition temperature $T^*$, which equals $T_c$ for $q\leq 4$, is highest equally at $q=3$ and $q=4$ among integer values of $q\geq 2$.
This optimality originates from the nature of the lower-order interactions \textit{induced by the simplicial interaction}, in the sense of Eq.~(\ref{eqn:Hamiltonian2}), within a hyperedge of size $q$.
The number ${{q}\choose{n}}$ of the subsets of $n$ spins, that contribute to the $n$-th order interaction energy represented by the $m^n$ term of $f(m)$, increases with $q$. In contrast, its effective interaction strength ${J_q\over 2^{q-1}}$ decreases with $q$.
Solving $C_2=0$, we obtain the transition temperature $T^*$, at which the local stability of $f(m)$ at $m=0$ becomes marginal, as
\begin{align}
    T^{*} = \frac{q (q - 1)}{2^{q - 1}},
\end{align}
which attains its maximum value ${3\over 2}$  at $q=3$ and $q=4$, while approaching zero in the limit $q\to\infty$.
Thus, groups of moderate size $3\le q\le5$ facilitate the transition from a disordered to an ordered state compared to pairwise interactions; however, too large a group $(q\ge6)$ would rather impede it.

\begin{figure}[t]
    \centering
    \includegraphics[width=.99\linewidth]{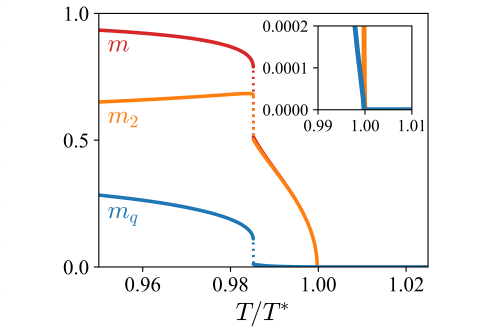}
    \caption{
    Decomposition of the magnetization order in a double transition. Using the Bethe--Peierls method, we plot the order parameter $m$ and its decomposed components $m_2$ and $m_q$, defined in Eq.~(\ref{eqn:contribution}), as functions of $T$ for $q=16$ and $(J_q, J_2, k_q, k_2)=(0.3,0.7,4,4)$. The inset shows a close-up near the continuous transition point, highlighting the singular behaviors of both $m_2$ and $m_q$.
    }
    \label{fig4}
\end{figure}

\subsection{Coexistence of pairs and groups}
Given the distinct nature of phase transitions---continuous for $q\le 4$ and discontinuous for $q>4$---on $q$-uniform hypergraphs, a natural question arises: What happens when hyperedge sizes are not uniform? Narrowing the focus, we consider hypergraphs where hyperedges are of size either $2$ (pairwise interaction) or $q>2$ (HOI), which we denote by $(2,q)$-hypergraphs. The propensity of HOIs is characterized  by a control parameter $r\in (0,1)$ such that  $\rho_{q} J_{q} = r$ and $\rho_{2} J_{2} = 1-r$.

Using Eq.~(\ref{eqn:free_energy}), we have obtained Fig.~\ref{fig3}(a), in which each point on the $(q,r)$-space with  $q>4$ represents a scenario of phase transition occurring as $T$ varies.
Intuitively, when $r$ is near $0$ or $1$, a continuous (Regime I) or a discontinuous (Regime II) transition, respectively, is expected to occur since either pairwise interactions or HOIs dominate the transition nature; indeed, the two regimes exist, divided by the TP line for $q \le 8$. However, this TP line terminates at the ``special'' TP, $(q,r,T)=(8,\frac{8}{57},\frac{35}{38})$, analytically obtained from $C_2=C_4=C_6=0$, while $C_8 > 0$, at which the critical exponent $\beta$ is $\beta_{\rm STP}={1 \over 6}$ in contrast to $\beta_{\rm TP} = {1 \over 4}$ of the other TPs.

Beyond the special TP, i.e., $q>8$, a new regime, Regime III, appears when the propensity of HOIs is moderate, i.e., for $r_{\text{CP}} < r < r_{\text{CE}}$, where a double transition occurs, consisting of a continuous transition followed by a discontinuous one as $T$ decreases.
Here, CP (CE) indicates the critical point (critical endpoint) line, dividing Regime I (Regime II) and Regime III, which we will explain below in detail. Figures~\ref{fig3}(b-d) show the phase transitions for $r=r_{\text{CP}}$, $r_{\text{CP}} < r < r_{\text{CE}}$, and $r=r_{\text{CE}}$ when $q=12$ in the $(2,q)$-hypergraphs.

As $r$ decreases at fixed $q$ within Regime III, the discontinuous jump at nonzero $m$ decreases and eventually vanishes at $r=r_{\text{CP}}$, turning into a continuous transition with the singularity in the susceptibility $\chi=\frac{\partial m}{\partial H}|_{H=0}$ on both sides [Fig.~\ref{fig3}(b)].
On the other hand, as $r$ increases, the temperatures for the continuous and discontinuous transitions approach each other and merge at $r=r_{\text{CE}}$.
Consequently, along the CE line, the system exhibits a single jump in $m$ at $T_c$, characteristic of a discontinuous transition, while retaining a feature of a continuous transition, the divergence of susceptibility as $T \to T_c^+$ [Fig.~\ref{fig3}(d)]. This defines the so-called mixed-order transition~\cite{bar2014mixed}.
Note that similar phase diagrams to Fig.~\ref{fig3}(a) have been reported from other generalized spin models~\cite{jang2015ashkin,kim2024entropy}.
Furthermore, the framework naturally extends to $(q_1, q_2)$-hypergraphs with both $q_1 > 4$ and $q_2 > q_1$, where double discontinuous transitions can arise; a systematic exploration of this generalization is left for future work.

\subsection{Multiscale nature of double transitions}
Double transitions in Regime III may provide key insights into the interplay between pairwise interactions and HOIs in establishing magnetization order in hypergraphs.
Moreover, given that pairwise interactions and HOIs lead to continuous transitions and discontinuous transitions, respectively, it is intriguing to examine the roles of these interactions and their interplay in giving rise to double transitions.
To investigate the mechanism of double transitions in $(2,q)$-hypergraphs, we quantify the contributions of pairwise interactions and HOIs to global magnetization using the Bethe--Peierls method~\cite{mezard2009information}. Here, we consider Bethe hyperlattices, where each node belongs to finite $k_2$ pairwise edges and $k_q$ size-$q$ hyperedges.

The Bethe--Peierls method allows us to introduce two effective fields (referred to as cavity biases~\cite{mezard2002analytic,mezard2002random}), $u_{2}$ and $u_{q}$ created by a hyperedge of size $2$ and $q$, respectively, at a spin~\cite{mezard2002random,mezard2002analytic,dorogovtsev2008critical}.
Specifically, $u_{q'}$, where $q'=2$ or $q$, for a central spin $S_0$ is defined as~\cite{mezard2009information}
\begin{equation}
\resizebox{0.95\columnwidth}{!}{$\displaystyle \sum_{S_1, \ldots, S_{q'-1}}e^{{\frac{1}{T}} \left[J_{q'}\delta_{\{S_0, S_1, \ldots, S_{q'-1}\}} + (h-u_{q'})\sum_{i=1}^{q'-1}{S_i}\right]}
    \propto e^{{\frac{1}{T}} u_{q'} S_{0}}$}
\end{equation}
with the total local field $h = k_{2} u_{2} + k_{q} u_{q}$ acting on each spin~(see Methods for details).
Using $u_2$ and $u_q$ obtained in a self-consistent way, one finds the order parameter given by $m = m_2 + m_q$ with
\begin{align}
\begin{split}
    m_{q'} = \frac{\tanh{\left(k_{q'} u_{q'}/T\right)} }{1+\tanh{\left(k_2 u_2 / T\right)}\tanh{\left( k_q u_q / T\right)}}
\end{split}
\label{eqn:contribution}
\end{align}
representing the contribution from pairwise interactions ($q'=2$) or HOIs ($q'=q$) to the global magnetization.

Figure~\ref{fig4} shows the results for $m$, $m_2$, and $m_q$ as  functions of $T$ from the Bethe--Peierls calculation for $q=16$ and $(J_q, J_2, k_q, k_2)=(0.3,0.7,4,4)$ as an instance of a double transition. Both $m_2$ and $m_q$ undergo both continuous and discontinuous transitions at the same temperatures, implying mutual influences of both types of interactions when they coexist; otherwise, $m_2$ would exhibit only a continuous transition and $m_q$ only a discontinuous transition.

However, the contributions of $m_2$ and $m_q$ to $m$ differ significantly in magnitude and temperature range, reflecting the multiscale nature of double transitions. In the intermediate ferromagnetic phase between the two transitions, we find $m_2\gg m_q$, indicating that pairwise interactions mainly drive the magnetization. Little contribution from HOIs is expected, given that the unanimous alignment of $q$ spins occurs only at significantly lower temperatures. Our analysis quantitatively supports this expectation.
Only at temperatures below the discontinuous transition temperature does the contribution of HOIs become comparable to that of pairwise interactions.
{In addition, $m_2$ decreases as $T$ decreases, indicating the relative contribution of groups to $m$ grows much more rapidly.}
Despite the smaller magnitude of $m_q$ compared to $m_2$, the role of HOIs is crucial, driving a discontinuous transition in the global magnetization.
These results indicate that the double transition originates from the distinct roles of interaction sizes across temperature.

\section{Conclusions}
To sum up, we have studied the simplicial Ising model (SIM) as a prototypical higher-order interaction (HOI) system.
We obtained rich phase diagrams on $q$-uniform and $(2,q)$-hypergraphs, revealing that the many-body unanimity interactions can give rise to a cornucopia of phase transition scenarios---including continuous, discontinuous, mixed-order, and double transitions---even without introducing complex structures.
We have shown that such diverse scenarios arise due to the effective lower-order couplings in the Hamiltonian induced by the simplicial HOIs, which lead to nontrivial lower-order corrections to the Landau free energy.
Moreover, such group interactions exhibit an ambivalent effect on the transition temperatures, leading to their optimality.
Finally, we characterized the multiscale nature of magnetic ordering in the SIM on $(2,q)$-hypergraphs by delineating the distinct contributions of pairwise interactions and HOIs in double transitions.

While we were completing this manuscript, we became aware of another group working independently on the same subject~\cite{robiglio2025higher}. Some results of that paper partly overlap with ours, such as the finding that phase transitions become more abrupt as the group size increases. {Beyond this overlap, the two works are complementary. Ref.~\cite{robiglio2025higher} compares the conserved-symmetry model (equivalent to the SIM) and the broken-symmetry model (traditional ferromagnetic $p$-spin models), finding their different critical behaviors. On the other hand, we focus on the SIM, revealing the conditions for and the multiscale nature of multiple transitions. To incorporate sparse connectivity, Ref.~\cite{robiglio2025higher} employs the Georges–Yedidia high-temperature expansion, while we apply the Bethe--Peierls method.}

Our results provide a cornerstone for further works.
First, investigating the SIM on hypergraphs with arbitrary degree and cardinality distributions could deepen our understanding.
For example, degree heterogeneity is expected to soften discontinuous transitions, as hubs sustain local order, while size heterogeneity may counteract this tendency; a detailed analysis of these competing effects will be reported elsewhere.
Second, by relaxing the unanimity condition, one can consider a generalized model where non-zero weights are assigned to all possible spin states within a hyperedge and explore how changes in the weight distribution affect the phase diagram.
In addition, the present framework can be directly extended beyond $\mathbb{Z}_2$ symmetry, e.g., the Potts model~\cite{wu1982potts} and $O(n)$-model~\cite{stanley1968dependence}.
From the perspective of social dynamics~\cite{castellano2009statistical,budrikis2024100}, our results predict that the abrupt transition is governed by a characteristic group size, which may inspire empirical studies in abrupt social change~\cite{centola2018experimental,iacopini2022group}.
Finally, recent advances in quantum technology have enabled controlled manipulation of multi-body interactions~\cite{katz2023demonstration} that could realize a quantum version of SIM.

\section{Methods}

\subsection{Mean-field methods}

In this section, we provide detailed descriptions of two mean-field methods employed in the main text.

\subsubsection{Bragg--Williams method}

We can rewrite the Hamiltonian in Eq.~(\ref{eqn:Hamiltonian}) as the following:
\begin{align}
    \mathcal{H}
    &= -\sum_{ q \ge 2 } J_{q} \left[ N_{+}^{(q)} + N_{-}^{(q)} \right] - H(N_+ - N_-)
\end{align}
where $N_{+}^{(q)}$ and $N_{-}^{(q)}$ denote the number of $q$-hyperedges in which all spins are $+1$ and $-1$, respectively, and $N_\pm$ are the number of $\pm1$ spins.
The Bragg--Williams method assumes that the spins are statistically independent. Then, $N_{\pm}$ can be written as $N_{\pm} = N(1 \pm m) / 2$ with the magnetization $m$. The entropy density $s$ is given by
\begin{align}
s=-\left(\frac{N_{+}}{N} \ln{\frac{N_{+}}{N}} + \frac{N_{-}}{N} \ln{\frac{N_{-}}{N}}\right),
\end{align}
and the number of unanimous hyperedges
\begin{align}
    \label{eqn:BW approx}
    N_{\pm}^{(q)} = |\mathcal{E}_q| \left(\frac{N_{\pm}}{N}\right)^{q}.
\end{align}
Finally, the free energy density in Eq.~(\ref{eqn:free_energy}) is derived.

\subsubsection{Bethe--Peierls method}
The Bethe--Peierls method retains the interactions of a spin with its neighbors, while encoding the coupling of each neighbor to the rest of the system as its external effective field~\cite{dorogovtsev2008critical,mezard2009information}.
Given a center spin $S_0$ and its $q$-hyperedge $\alpha=\{0, 1,\ldots, {q-1}\}$, we introduce a ``partial" partition function $Z_{\alpha \to 0}(S_0)$ by summing over all spins in $\alpha$ except $S_0$:
\begin{align}
\begin{split}
Z_{\alpha \to 0}(S_0) &= \sum_{S_1 \ldots S_{q-1}} e^{J\delta_{\{S_i\}_{i=0}^{q-1}} + \sum_{i=1}^{q-1} h_i S_i}\\
&\equiv c(J,\{h_i\})e^{u(J,\{h\})S_0}
\end{split}\\
\begin{split}
&= e^{J\delta_{S_0,1} + \sum_{i=1}^{q-1} h_i} + e^{J\delta_{S_0,-1} - \sum_{i=1}^{q-1} h_i}\\
&\quad+ \underbrace{\sum_{\substack{S_1 \ldots S_{q-1}\\ \delta_{\{S_i\}_{i=1}^{q-1}} = 0}} e^{\sum_{i=1}^{q-1} h_i S_i}}_{\mathcal{X}}
\end{split}\\
&= 2 e^{\frac{J}{2}}\cosh{\left( \frac{J}{2}S_0 +\sum_{i=1}^{q-1} h_i \right)} + \mathcal{X}
\end{align}

where we used the simplified notations for the cavity field $h_i \equiv h_{i\to\alpha} = \sum_{\alpha' \in \partial i \setminus \alpha} u_{\alpha' \to i}$, the cavity bias upon the central spin $u \equiv u_{\alpha \to 0}$, and the ferromagnetic coupling constant $J \equiv J_q$, with $\partial i$ being the set of hyperedges incident to node $i$. Finally, the cavity bias $u_{\alpha \to 0} \equiv \frac{1}{2} \log{\frac{Z_{\alpha \to 0}(+1)}{Z_{\alpha \to 0}(-1)}}$, can be written as
\begin{align}
    \label{eqn:cavity_bias}
    u_{\alpha \to 0}
    = \tanh^{-1}{\left[ \frac{\mathcal{S}(\{h_{i\to\alpha}\}_{i\in\alpha'})}{\mathcal{C}(\{h_{i\to\alpha}\}_{i\in\alpha'}) + \mathcal{X}(\{h_{i\to\alpha}\}_{i\in\alpha'})} \right]}
\end{align}
where
\begin{align}
    \mathcal{S}(\{h_{i\to\alpha}\}_{i\in\alpha'}) &= 2 e^{J/2} \sinh{(J/2)} \sinh{\left(\sum_{i\in\alpha'} h_{i\to\alpha}\right)}\\
    \mathcal{C}(\{h_{i\to\alpha}\}_{i\in\alpha'}) &= 2 e^{J/2} \cosh{(J/2)} \cosh{\left(\sum_{i\in\alpha'} h_{i\to\alpha}\right)}\\
    \mathcal{X}(\{h_{i\to\alpha}\}_{i\in\alpha'}) &= \prod_{i\in\alpha'} \left. 2 \cosh{h_{i\to\alpha}} \right. - 2 \cosh{\left( \sum_{i\in\alpha'} h_{i\to\alpha} \right)},
\end{align}
and $\alpha'\equiv\alpha\setminus\{0\}=\{1,\ldots, {q-1}\}$.
Note that the omitted temperature $T$ is recovered by $u\to u/T$, $J\to J/T$, and $h\to h/T$.

In the main text, we consider Bethe hyperlattices in which each node is incident to $k_2$ pairwise edges and $k_q$ size-$q$ hyperedges to quantify the contribution of each interaction size in the double transitions. While the cavity bias $u_{\alpha \to i}$ is assigned for each hyperedge--node pair $(\alpha,i)$, in the regular Bethe hyperlattices we considered, the bias depends only on the size of $\alpha$. Consequently, Eq.~(\ref{eqn:cavity_bias}) yields self-consistent equations in terms of only two cavity bias variables, $u_2$ for pairwise edges and $u_q$ for $q$-hyperedges, as follows:
\begin{align}
    \tanh{u_{2}} &= \tanh{(J_{2}/2)} \tanh{(h-u_{2})}\\
    \tanh{u_{q}} &=
    \frac
    {(e^{J_{q}}-1) \sinh{[(q - 1)(h - u_{q})]}}
    {\begin{gathered}(e^{J_{q}}-1) \cosh{[(q - 1)(h - u_{q})]}\\ {}+ [2\cosh{(h - u_{q})}]^{q - 1}\end{gathered}}
\end{align}
where $h = k_2 u_2 + k_q u_q$.

\subsection{Asymptotic behavior of transition temperatures in uniform hypergraphs}
Among the three characteristic temperatures, $T^*$, $T_c$, and $T^{**}$, in the SIM on $q$-uniform hypergraphs shown in Fig.~\ref{fig2}, $T^*$ is amenable to a closed-form solution, since the shape of the free energy only near $m=0$ governs it. For the others, $T^{**}$ and $T_c$, we obtain their asymptotic behaviors when $q$ is large enough.
The intuitive picture under this investigation is the fact that for large $q$, the non-zero minima are formed near $m=\pm1$, since $\left( \frac{1 \pm m}{2} \right)^{q} \approx \delta_{m,\pm1}$.
The asymptotic of $T_c$ can be obtained from the temperature at which $f(1)=f(0)$; thus, $T_c \to \frac{1}{\ln{2}}=1.442695\ldots$.
Furthermore, the large-$q$ asymptotic of $T^{**}$ can be obtained by expanding the free energy density near $|m|=1$. From the conditions $f'(m)=0$ and $f''(m)=0$ at $m=m^{**} \equiv 1-\epsilon$ with $\epsilon \ll 1$ and $T=T^{**}$,
we find $T^{**} \sim - \frac{q}{\ln{\epsilon}} \approx - \frac{q}{W_{-1}(-1/q)}$ where $W_{-1}(x) \approx \ln{(-x)}-\ln{(-\ln{(-x)})}$ is the secondary branch of the Lambert $W$-function, i.e., the inverse of $W e^{W}$.

\subsection{Monte Carlo simulation}
To conduct the Monte Carlo (MC) simulation in Fig.~2(b) of the main text, we adopt the Metropolis algorithm. Specifically, for each step $\tau$, we randomly select a spin $S_i$ and accept its flip with probability:
\begin{align}
    \min{\left[1,e^{-\Delta E_i (\tau) / T}\right]}
\label{eqn:metropolis}
\end{align}
where $\Delta E_i (\tau) = \mathcal{H}(\{S_j(\tau+1)\}) - \mathcal{H}(\{S_j(\tau)\})$ and $S_j(\tau+1)=S_j(\tau)$ for all $j \neq i$, while $S_i(\tau+1)=-S_i(\tau)$.
We perform simulations on fully connected $10$-uniform hypergraphs with $N=10^4$. Each run consists of $10^4$ MC steps for equilibration and $2\times10^4$ MC steps for measuring $m$, and the results are further averaged over 10 runs with different initial configurations.

\section*{Acknowledgements}

This work was supported by a KIAS Individual Grant [No. CG098001 (GS) and No. CG079902 (D-SL)] at Korea Institute for Advanced Study; and by National Research Foundation of Korea (NRF) grants funded by the Korean government (MSIT) [No. 2020R1A2C2003669 and No. RS-2025-00558837] (K-IG).

\bibliography{refs}

@article{boccaletti2023structure,
  title={The structure and dynamics of networks with higher order interactions},
  author={Boccaletti, Stefano and De Lellis, Pietro and Del Genio, CI and Alfaro-Bittner, Karin and Criado, Regino and Jalan, Sarika and Romance, Miguel},
  journal={Physics Reports},
  volume={1018},
  pages={1--64},
  year={2023},
  publisher={Elsevier}
}

@article{battiston2020networks,
  title={Networks beyond pairwise interactions: Structure and dynamics},
  author={Battiston, Federico and Cencetti, Giulia and Iacopini, Iacopo and Latora, Vito and Lucas, Maxime and Patania, Alice and Young, Jean-Gabriel and Petri, Giovanni},
  journal={Physics Reports},
  volume={874},
  pages={1--92},
  year={2020},
  publisher={Elsevier}
}

@book{bianconi2021higher,
  title={Higher-order networks},
  author={Bianconi, Ginestra},
  year={2021},
  publisher={Cambridge University Press}
}

@article{battiston2021physics,
  title={The physics of higher-order interactions in complex systems},
  author={Battiston, Federico and Amico, Enrico and Barrat, Alain and Bianconi, Ginestra and Ferraz de Arruda, Guilherme and Franceschiello, Benedetta and Iacopini, Iacopo and K{\'e}fi, Sonia and Latora, Vito and Moreno, Yamir and others},
  journal={Nature Physics},
  volume={17},
  number={10},
  pages={1093--1098},
  year={2021},
  publisher={Nature Publishing Group UK London}
}

@article{robiglio2025synergistic,
  title={Synergistic signatures of group mechanisms in higher-order systems},
  author={Robiglio, Thomas and Neri, Matteo and Coppes, Davide and Agostinelli, Cosimo and Battiston, Federico and Lucas, Maxime and Petri, Giovanni},
  journal={Physical review letters},
  volume={134},
  number={13},
  pages={137401},
  year={2025},
  publisher={APS}
}

@article{iacopini2019simplicial,
  title={Simplicial models of social contagion},
  author={Iacopini, Iacopo and Petri, Giovanni and Barrat, Alain and Latora, Vito},
  journal={Nature Communications},
  volume={10},
  number={1},
  pages={2485},
  year={2019},
  publisher={Nature Publishing Group UK London}
}

@article{kuehn2021universal,
  title={A universal route to explosive phenomena},
  author={Kuehn, Christian and Bick, Christian},
  journal={Science Advances},
  volume={7},
  number={16},
  pages={eabe3824},
  year={2021},
  publisher={American Association for the Advancement of Science}
}

@article{bianconi2024theory,
  title={Theory of percolation on hypergraphs},
  author={Bianconi, Ginestra and Dorogovtsev, Sergey N},
  journal={Physical Review E},
  volume={109},
  number={1},
  pages={014306},
  year={2024},
  publisher={APS}
}

@article{bianconi2024nature,
  title={Nature of hypergraph k-core percolation problems},
  author={Bianconi, Ginestra and Dorogovtsev, Sergey N},
  journal={Physical Review E},
  volume={109},
  number={1},
  pages={014307},
  year={2024},
  publisher={APS}
}

@article{lee2023k,
  title={(k, q)-core decomposition of hypergraphs},
  author={Lee, Jongshin and Goh, Kwang-Il and Lee, Deok-Sun and Kahng, B},
  journal={Chaos, Solitons \& Fractals},
  volume={173},
  pages={113645},
  year={2023},
  publisher={Elsevier}
}

@article{kim2024higher,
  title={Higher-Order Components Dictate Higher-Order Contagion Dynamics in Hypergraphs},
  author={Kim, Jung-Ho and Goh, K-I},
  journal={Physical Review Letters},
  volume={132},
  number={8},
  pages={087401},
  year={2024},
  publisher={APS}
}

@article{kim2025competition,
  title={Competition between group interactions and nonlinearity in voter dynamics on hypergraphs},
  author={Kim, Jihye and Lee, Deok-Sun and Min, Byungjoon and Porter, Mason A and San Miguel, Maxi and Goh, K-I},
  journal={Physical Review E},
  volume={111},
  number={5},
  pages={L052301},
  year={2025},
  publisher={APS}
}

@article{horstmeyer2020adaptive,
  title={Adaptive voter model on simplicial complexes},
  author={Horstmeyer, Leonhard and Kuehn, Christian},
  journal={Physical Review E},
  volume={101},
  number={2},
  pages={022305},
  year={2020},
  publisher={APS}
}

@article{papanikolaou2022consensus,
  title={Consensus from group interactions: An adaptive voter model on hypergraphs},
  author={Papanikolaou, Nikos and Vaccario, Giacomo and Hormann, Erik and Lambiotte, Renaud and Schweitzer, Frank},
  journal={Physical Review E},
  volume={105},
  number={5},
  pages={054307},
  year={2022},
  publisher={APS}
}

@article{millan2020explosive,
  title={Explosive higher-order Kuramoto dynamics on simplicial complexes},
  author={Mill{\'a}n, Ana P and Torres, Joaqu{\'\i}n J and Bianconi, Ginestra},
  journal={Physical Review Letters},
  volume={124},
  number={21},
  pages={218301},
  year={2020},
  publisher={APS}
}

@article{bar2014mixed,
  title={Mixed-order phase transition in a one-dimensional model},
  author={Bar, Amir and Mukamel, David},
  journal={Physical Review Letters},
  volume={112},
  number={1},
  pages={015701},
  year={2014},
  publisher={APS}
}

@article{jang2015ashkin,
  title={Ashkin-Teller model and diverse opinion phase transitions on multiplex networks},
  author={Jang, S and Lee, JS and Hwang, S and Kahng, B},
  journal={Physical Review E},
  volume={92},
  number={2},
  pages={022110},
  year={2015},
  publisher={APS}
}

@article{kim2024entropy,
  title={Entropy-induced phase transitions in a hidden Potts model},
  author={Kim, Cook Hyun and Lee, D-S and Kahng, Byungnam},
  journal={Physical Review E},
  volume={110},
  number={2},
  pages={024133},
  year={2024},
  publisher={APS}
}

@article{dorogovtsev2002ising,
  title={Ising model on networks with an arbitrary distribution of connections},
  author={Dorogovtsev, Sergey N and Goltsev, Alexander V and Mendes, Jos{\'e} Fernando F},
  journal={Physical Review E},
  volume={66},
  number={1},
  pages={016104},
  year={2002},
  publisher={APS}
}

@article{dorogovtsev2008critical,
  title={Critical phenomena in complex networks},
  author={Dorogovtsev, Sergey N and Goltsev, Alexander V and Mendes, Jos{\'e} FF},
  journal={Reviews of Modern Physics},
  volume={80},
  number={4},
  pages={1275--1335},
  year={2008},
  publisher={APS}
}

@article{leone2002ferromagnetic,
  title={Ferromagnetic ordering in graphs with arbitrary degree distribution},
  author={Leone, M and V{\'a}zquez, A and Vespignani, A and Zecchina, Riccardo},
  journal={The European Physical Journal B},
  volume={28},
  pages={191--197},
  year={2002},
  publisher={Springer}
}

@article{bianconi2002mean,
  title={Mean field solution of the Ising model on a Barab{\'a}si--Albert network},
  author={Bianconi, Ginestra},
  journal={Physics Letters A},
  volume={303},
  number={2-3},
  pages={166--168},
  year={2002},
  publisher={Elsevier}
}

@article{gross1984simplest,
  title={The simplest spin glass},
  author={Gross, David J and M{\'e}zard, Marc},
  journal={Nuclear Physics B},
  volume={240},
  number={4},
  pages={431--452},
  year={1984},
  publisher={Elsevier}
}

@article{di2024dynamical,
  title={Dynamical Fluctuations of Random Walks in Higher-Order Networks},
  author={Di Gaetano, Leonardo and Carugno, Giorgio and Battiston, Federico and Coghi, Francesco},
  journal={Physical Review Letters},
  volume={133},
  number={10},
  pages={107401},
  year={2024},
  publisher={APS}
}

@article{civilini2021evolutionary,
  title={Evolutionary game model of group choice dilemmas on hypergraphs},
  author={Civilini, Andrea and Anbarci, Nejat and Latora, Vito},
  journal={Physical Review Letters},
  volume={127},
  number={26},
  pages={268301},
  year={2021},
  publisher={APS}
}

@article{civilini2024explosive,
  title={Explosive cooperation in social dilemmas on higher-order networks},
  author={Civilini, Andrea and Sadekar, Onkar and Battiston, Federico and G{\'o}mez-Garde{\~n}es, Jes{\'u}s and Latora, Vito},
  journal={Physical Review Letters},
  volume={132},
  number={16},
  pages={167401},
  year={2024},
  publisher={APS}
}

@article{burgio2024triadic,
  title={Triadic approximation reveals the role of interaction overlap on the spread of complex contagions on higher-order networks},
  author={Burgio, Giulio and G{\'o}mez, Sergio and Arenas, Alex},
  journal={Physical Review Letters},
  volume={132},
  number={7},
  pages={077401},
  year={2024},
  publisher={APS}
}

@article{kim2023contagion,
  title={Contagion dynamics on hypergraphs with nested hyperedges},
  author={Kim, Jihye and Lee, Deok-Sun and Goh, K-I},
  journal={Physical Review E},
  volume={108},
  number={3},
  pages={034313},
  year={2023},
  publisher={APS}
}

@article{carletti2020random,
  title={Random walks on hypergraphs},
  author={Carletti, Timoteo and Battiston, Federico and Cencetti, Giulia and Fanelli, Duccio},
  journal={Physical Review E},
  volume={101},
  number={2},
  pages={022308},
  year={2020},
  publisher={APS}
}

@article{skardal2019abrupt,
  title={Abrupt desynchronization and extensive multistability in globally coupled oscillator simplexes},
  author={Skardal, Per Sebastian and Arenas, Alex},
  journal={Physical Review Letters},
  volume={122},
  number={24},
  pages={248301},
  year={2019},
  publisher={APS}
}

@article{st2021universal,
  title={Universal nonlinear infection kernel from heterogeneous exposure on higher-order networks},
  author={St-Onge, Guillaume and Sun, Hanlin and Allard, Antoine and H{\'e}bert-Dufresne, Laurent and Bianconi, Ginestra},
  journal={Physical Review Letters},
  volume={127},
  number={15},
  pages={158301},
  year={2021},
  publisher={APS}
}

@article{ferraz2024contagion,
  title={Contagion dynamics on higher-order networks},
  author={Ferraz de Arruda, Guilherme and Aleta, Alberto and Moreno, Yamir},
  journal={Nature Reviews Physics},
  volume={6},
  number={8},
  pages={468--482},
  year={2024},
  publisher={Nature Publishing Group UK London}
}

@article{noonan2021dynamics,
  title={Dynamics of majority rule on hypergraphs},
  author={Noonan, James and Lambiotte, Renaud},
  journal={Physical Review E},
  volume={104},
  number={2},
  pages={024316},
  year={2021},
  publisher={APS}
}

@article{franz2001ferromagnet,
  title={A ferromagnet with a glass transition},
  author={Franz, Silvio and M{\'e}zard, Marc and Ricci-Tersenghi, Federico and Weigt, Martin and Zecchina, Riccardo},
  journal={Europhysics Letters},
  volume={55},
  number={4},
  pages={465},
  year={2001},
  publisher={IOP Publishing}
}

@book{mezard2009information,
  title={Information, physics, and computation},
  author={Mezard, Marc and Montanari, Andrea},
  year={2009},
  publisher={Oxford University Press}
}

@article{edwards1975theory,
  title={Theory of spin glasses},
  author={Edwards, Samuel Frederick and Anderson, Phil W},
  journal={Journal of Physics F: Metal Physics},
  volume={5},
  number={5},
  pages={965},
  year={1975},
  publisher={IOP Publishing}
}

@article{solon2013revisiting,
  title={Revisiting the flocking transition using active spins},
  author={Solon, Alexandre P and Tailleur, Julien},
  journal={Physical review letters},
  volume={111},
  number={7},
  pages={078101},
  year={2013},
  publisher={APS}
}

@article{wu1982potts,
  title={The potts model},
  author={Wu, Fa-Yueh},
  journal={Reviews of modern physics},
  volume={54},
  number={1},
  pages={235},
  year={1982},
  publisher={APS}
}

@article{stanley1968dependence,
  title={Dependence of critical properties on dimensionality of spins},
  author={Stanley, H Eugene},
  journal={Physical Review Letters},
  volume={20},
  number={12},
  pages={589},
  year={1968},
  publisher={APS}
}

@article{budrikis2024100,
  title={100 years of the Ising model},
  author={Budrikis, Zoe},
  journal={Nature Reviews Physics},
  volume={6},
  number={9},
  pages={530--530},
  year={2024},
  publisher={Nature Publishing Group UK London}
}

@article{castellano2009statistical,
  title={Statistical physics of social dynamics},
  author={Castellano, Claudio and Fortunato, Santo and Loreto, Vittorio},
  journal={Reviews of modern physics},
  volume={81},
  number={2},
  pages={591--646},
  year={2009},
  publisher={APS}
}

@article{katz2023demonstration,
  title={Demonstration of three-and four-body interactions between trapped-ion spins},
  author={Katz, Or and Feng, Lei and Risinger, Andrew and Monroe, Christopher and Cetina, Marko},
  journal={Nature Physics},
  volume={19},
  number={10},
  pages={1452--1458},
  year={2023},
  publisher={Nature Publishing Group UK London}
}

@article{stauffer2008social,
  title={Social applications of two-dimensional Ising models},
  author={Stauffer, Dietrich},
  journal={American Journal of Physics},
  volume={76},
  number={4},
  pages={470--473},
  year={2008},
  publisher={AIP Publishing}
}

@article{lee2015statistical,
  title={Statistical mechanics of the US Supreme Court},
  author={Lee, Edward D and Broedersz, Chase P and Bialek, William},
  journal={Journal of Statistical Physics},
  volume={160},
  number={2},
  pages={275--301},
  year={2015},
  publisher={Springer}
}

@article{mezard2002analytic,
  title={Analytic and algorithmic solution of random satisfiability problems},
  author={M{\'e}zard, Marc and Parisi, Giorgio and Zecchina, Riccardo},
  journal={Science},
  volume={297},
  number={5582},
  pages={812--815},
  year={2002},
  publisher={American Association for the Advancement of Science}
}

@article{mezard2002random,
  title={Random k-satisfiability problem: From an analytic solution to an efficient algorithm},
  author={M{\'e}zard, Marc and Zecchina, Riccardo},
  journal={Physical Review E},
  volume={66},
  number={5},
  pages={056126},
  year={2002},
  publisher={APS}
}

@article{robiglio2025higher,
  title={Higher-order Ising model on hypergraphs},
  author={Robiglio, Thomas and Di Gaetano, Leonardo and Altieri, Ada and Petri, Giovanni and Battiston, Federico},
  journal={Physical Review E},
  volume={112},
  number={2},
  pages={L022301},
  year={2025},
  publisher={APS}
}

@article{centola2018experimental,
  title={Experimental evidence for tipping points in social convention},
  author={Centola, Damon and Becker, Joshua and Brackbill, Devon and Baronchelli, Andrea},
  journal={Science},
  volume={360},
  number={6393},
  pages={1116--1119},
  year={2018},
  publisher={American Association for the Advancement of Science}
}

@article{iacopini2022group,
  title={Group interactions modulate critical mass dynamics in social convention},
  author={Iacopini, Iacopo and Petri, Giovanni and Baronchelli, Andrea and Barrat, Alain},
  journal={Communications Physics},
  volume={5},
  number={1},
  pages={64},
  year={2022},
  publisher={Nature Publishing Group UK London}
}
\end{document}